\documentclass[12pt]{article}
\author{N.V. Ustinov\\
\em Quantum Field Theory Department, Tomsk State University,\\
\em 36 Lenin Avenue, Tomsk, 634050, Russia}
\title{Infinitesimal symmetries and conservation laws of the DNLSE hierarchy
and the Noether's theorem}
\date{ }
\begin{document}
\maketitle
\begin{abstract}
The hierarchy of the integrable nonlinear equations associated with the 
quadratic bundle is considered. 
The expressions for the solution of the linearization of these equations and 
their conservation law in the terms of the solutions of the corresponding Lax 
pairs are found. 
It is shown for the first member of the hierarchy that the conservation law is 
connected with the solution of the linearized equation due to the Noether's 
theorem. 
The local hierarchy and three nonlocal ones of the infinitesimal symmetries 
and the conservation laws that are explicitly expressed through the variables 
of the nonlinear equations are derived. 
\end{abstract}

\section{Introduction}
One of the most effective tools of studying the nonlinear phenomena is the 
inverse scattering transformation (IST) method \cite{NMPZ,AS}. 
This method reduces the solution of the Cauchy initial value problem of the 
nonlinear partial differential equations (PDE's), which admit a representation 
as the compatibility condition of the overdetermined linear system (Lax pair), 
to solving linear singular integral equations. 
It is of a special significance that many of the PDE's playing an important 
role in different branches of physics can be investigated in its frameworks. 
For example, the derivative nonlinear Schr\"odinger equation (DNLSE) that 
was originally deduced for the Alfv\'en waves of finite amplitude 
\cite{MOMT,M} and the equations of massive Thirring model (MTM) \cite{Th} 
belong to the class of the PDE's integrable with the help of the IST method 
for the quadratic bundle 
[6--8].
It was revealed that DNLSE describes also the behavior of drifting 
filamentations in nonlinear electrostatic waves of magnetized plasmas 
\cite{SShY}, light pulses in the optical fibers 
[10--12], 
magnetic holes of space plasmas \cite{B} and large-amplitude 
magnetohydrodynamics waves \cite{R}. 
The MTM equations were recently shown to appear in the coherent optics and 
nonlinear acoustics \cite{Z1} as a limiting case of the system of 
long/short-wave coupling (see \cite{SU} and references therein). 

The integrable nonlinear PDE's are well known to possess the infinite 
hierarchies of infinitesimal symmetries and conservation laws. 
An existence of them was proposed as the integrability test to characterize 
the equations solvable by IST (see, e.g., \cite{Fokas,MShS}). 
There are different methods of obtaining the infinitesimal symmetries and 
conservation laws, which originate from the study of KdV equation 
\cite{Miura}. 
Given a set of the scattering data (namely, the time-invariant part of them), 
infinite hierarchies of the conserved densities are constructed 
[1,\ 2,\ 6,\ 20--22].
The B\"acklund transformation (BT) of the integrable equation was used to 
generate the hierarchy of its conservation laws \cite{K,DB}. 
The approach that exploits the Noether's theorem was applied in 
\cite{St_1,St_2} for the derivation of the conservation laws of sine--Gordon 
and KdV equations. 
To produce the corresponding hierarchy of infinitesimal symmetries, 
the implicit expressions for the solutions of the linearized equations, which 
are obtained by means of infinitesimal BT, were expanded in the power series 
on the parameter of this BT. 
The infinitesimal version of the dressing method was suggested in \cite{OSch} 
to construct the infinitesimal symmetries of integrable PDE's. 
Similar expressions for the perturbations of some nonlinear PDE's and their 
Lax pairs were presented in \cite{S}. 
The geometrical approaches that utilize the projective transformations or 
treat the soliton equations as descriptions of pseudospherical surfaces were 
developed for nonlinear PDE's associated with matrix Lax pairs of the second 
order in \cite{S_1,S_2} (see also \cite{KSI}) and \cite{CT}, respectively. 
The hierarchies of local and nonlocal conservation laws for DNLSE were 
found by means of these methods \cite{S_2,W}. 
The method based on the theory of $\tau$-functions was applied to scalar and 
two-component KP hierarchies \cite{MaSS}. 

Although the methods mentioned above appeal to underlying Lax pair to produce 
the hierarchies of infinitesimal symmetries and conservation laws, they do 
not entirely cover the class of PDE's representable as the compatibility 
condition. 
This concerns especially the cases of reductions of the nonlinear PDE's 
\cite{Mikh} and their integrable deformations (see, e.g., \cite{BZ,Z1,Z2}), 
which are most interesting from physical point of view. 
The knowledge of the infinitesimal symmetries and conservation densities 
of the hierarchy allows one to make sure that the PDE given belongs to it. 
The approach applicable to all integrable nonlinear equations can be based, 
for instance, on explicit expressions for the solution of linearized equation 
and the conservation law in the terms of the solutions of corresponding Lax 
pairs. 
In the present report, we construct the infinite hierarchies of local and 
nonlocal infinitesimal symmetries and conservation laws for the DNLSE 
hierarchy using such the approach. 

The paper is organized as follows.
The nonlinear equations of the DNLSE hierarchy and their Lax pairs are 
presented in Sec.II.
The formulas for the expansions in series on the spectral parameter powers 
of the solutions of the Lax pairs are also given there. 
The solution of the linearization of the nonlinear PDE's, which is expressed 
in the terms of the solutions of corresponding Lax pairs, is obtained by means 
of the infinitesimal version of the binary Darboux transformation (DT) 
\cite{MS} in Sec.III. 
This technique has been applied to the DNLSE 
[38--40] and to obtain the infinitesimal symmetries of the nonlinear PDE's 
[41--43]. 
To generate the hierarchies of local and nonlocal infinitesimal symmetries 
explicitly expressed through the variables of the nonlinear equations under 
consideration, the expansions in series of the solutions of the Lax pairs or 
the recursion operator of the hierarchy can be used.
In Sec.IV the conservation law for the DNLSE hierarchy is derived.
Substitution of the expansion in series of the Lax pairs solutions into this 
formula yields the hierarchies of local and nonlocal conservation laws. 
The connection due to the Noether's theorem of the infinitesimal symmetries 
and conservation laws we found is shown in this section for the DNLSE case.

\section{The DNLSE hierarchy}
Let us consider (direct) Lax pair
\begin{equation}
\psi_x=U(\lambda)\,\psi,
\label{psi_x}
\end{equation}
\begin{equation}
\psi_t=V(\lambda)\,\psi,
\label{psi_t}
\end{equation}
where $\psi=\psi(x,t,\lambda)=(\psi_1,\psi_2)^T$ is the vector-column 
solution; $\lambda$ is complex parameter referred to as the spectral parameter 
in the IST theory; $U(\lambda)=U(x,t,\lambda)$ and $V(\lambda)=V(x,t,\lambda)$ 
are $2\times2$ matrix coefficients.
The compatibility condition of the overdetermined system 
(\ref{psi_x},\ref{psi_t}) is 
\begin{equation}
U(\lambda)_t-V(\lambda)_x+[\,U(\lambda)\,,V(\lambda)\,]=0.
\label{cc}
\end{equation}

We suppose in what follows that 
\begin{equation}
U(\lambda)=\lambda^2U^{(2)}+\lambda\,U^{(1)}
\label{U}
\end{equation}
(i.e., Eq.(\ref{psi_x}) is the quadratic bundle) and 
\begin{equation}
U^{(2)}=
\left(
\begin{array}{cc}
-i&0\\
0&i
\end{array}
\right),\quad
U^{(1)}=
\left(
\begin{array}{cc}
0&q\\
r&0
\end{array}
\right). 
\label{U's}
\end{equation}
If $V(\lambda)$ is chosen in the next form 
\begin{equation}
V(\lambda)=\sum_{j=1}^{2m}\lambda^jV^{(j)},
\label{V}
\end{equation}
then Eq.(\ref{cc}) gives the expressions for the matrix coefficients of 
$V(\lambda)$:
\begin{equation}
V^{(2m-2j)}=v^{(2m-2j)}
\left(
\begin{array}{cc}
1&0\\
0&-1
\end{array}
\right),\quad
V^{(2m-2j-1)}=
\left(
\begin{array}{cc}
0&v^{(2m-2j-1)}_{12}\\
v^{(2m-2j-1)}_{21}&0
\end{array}
\right),
\label{V's}
\end{equation}
($j=0,...,m-1$), where
\begin{equation}
v^{(2m-2j)}=\partial_x^{-1}(qv^{(2m-2j-1)}_{21}-rv^{(2m-2j-1)}_{12})\,,\quad
\label{v's1}
\end{equation}
\begin{equation}
\left(
\begin{array}{c}
{v^{(2m-2j-1)}_{12}}_{\mathstrut}\\
{v^{(2m-2j-1)}_{21}}^{\mathstrut}
\end{array}
\right)=\widehat{R}^{j+1}\left(
\begin{array}{c}
{0_{\mathstrut}}_{\mathstrut}\\
{0^{\mathstrut}}^{\mathstrut}
\end{array}
\right),
\label{v's2}
\end{equation}
\begin{equation}
\widehat{R}=\frac12
\left(
\begin{array}{cc}
i\partial_x+q\partial_x^{-1}r\partial_x&q\partial_x^{-1}q\partial_x\\
r\partial_x^{-1}r\partial_x&-i\partial_x+r\partial_x^{-1}q\partial_x
\end{array}
\right), 
\label{Rh}
\end{equation}
and system of nonlinear equations
\begin{equation}
\left(
\begin{array}{c}
q_t\\
r_t
\end{array}
\right)=\partial_x\widehat{R}^m\left(
\begin{array}{c}
0\\
0
\end{array}
\right). 
\label{h}
\end{equation}
(Note that operators $\partial_x^{-1}$ in Eqs.(\ref{v's1},\ref{v's2}) for 
equal $j$'s add the same time--dependent functions as the constants of 
integration.)
To obtain these formulas we make use the identities
$$
R\partial_x=\partial_x\widehat{R},
$$
\begin{equation}
R^{-1}=2
\left(
\begin{array}{cc}
-i+q\partial_x^{-1}r&-q\partial_x^{-1}q\\
-r\partial_x^{-1}r&i+r\partial_x^{-1}q
\end{array}
\right)
\left(
\begin{array}{cc}
\partial_x^{-1}&0\\
0&\partial_x^{-1}
\end{array}
\right)
\label{invR}
\end{equation}
with operator $R$ being defined in the following manner
\begin{equation}
R=\frac12
\left(
\begin{array}{cc}
i\partial_x+\partial_xq\partial_x^{-1}r&\partial_xq\partial_x^{-1}q\\
\partial_xr\partial_x^{-1}r&-i\partial_x+\partial_xr\partial_x^{-1}q
\end{array}
\right).
\label{R}
\end{equation}
As it will be seen in the next section, $\widehat{R}$ and its adjoint $R$ are 
the squared eigenfunction operator and the recursion one \cite{Fokas} of the 
hierarchy considered. 

The hierarchy of nonlinear equations (\ref{h}) was found in \cite{GIK}. 
It admits under appropriate choice of the constants of integration the next 
reduction 
\begin{equation}
r=\pm q^*. 
\label{rc}
\end{equation}
In this case, the first nontrivial equation of the hierarchy is reduced after 
rescaling to DNLSE 
\begin{equation}
iq_t+q_{xx}\mp i(|q|^2q)_x=0.
\label{DNLSE}
\end{equation}

Let us consider the expansions of the solutions of 
Eqs.(\ref{psi_x},\ref{psi_t}) in the series on the spectral parameter powers. 
In the neighborhood of point $\lambda=\infty$, the vector solutions of the Lax 
pairs of nonlinear equations (\ref{h}) are represented as 
\begin{equation}
\psi=\sum\limits_{k=0}^{\infty}\lambda^{-k}A^{(k)}\Lambda\,|a\rangle.
\label{psi_inf}
\end{equation}
Here $|a\rangle$ is a constant vector-column,
$$
\Lambda=\left(
\begin{array}{cc}
\mbox{e}^{\displaystyle-i\lambda^2x+\lambda^{2m}v^{(2m)}t}&0\\
0&\mbox{e}^{\displaystyle i\lambda^2x-\lambda^{2m}v^{(2m)}t}
\end{array}
\right)
$$
and coefficients $A^{(k)}$ solve system of equations
$$
\left\{
\begin{array}{l}
[A^{(k)},U^{(2)}]+A_x^{(k-2)}=U^{(1)}A^{(k-1)}\\
\mbox{}[A^{(k)},V^{(2m)}]+A_t^{(k-2m)}=
\sum\limits_{j=1}^{2m-1}V^{(2m-j)}A^{(k-j)}
\end{array}
\right..
$$
An expansion in series of the solutions of Lax pairs considered in the 
neighborhood of point $\lambda=0$ has form
\begin{equation}
\psi=\sum\limits_{k=0}^{\infty}\lambda^kB^{(k)}\,|a\rangle,
\label{psi_0}
\end{equation}
where $B^{(0)}=E$ and coefficients $B^{(k)}$ ($k\ge1$) are determined from
equations
$$
\left\{
\begin{array}{l}
B_x^{(k)}=U^{(2)}B^{(k-2)}+U^{(1)}B^{(k-1)}\\
B_t^{(k)}=\sum\limits_{j=1}^{2m}V^{(j)}B^{(k-j)}
\end{array}
\right..
$$
It is seen that the first coefficients of the expansions are 
$$
A^{(0)}=\left(
\begin{array}{cc}
w&0\\
0&w^{-1}
\end{array}
\right),\quad
A^{(1)}=\frac{i}{2}\left(
\begin{array}{cc}
0&-qw^{-1}\\
rw&0
\end{array}
\right),
$$
$$
A^{(2)}=\frac14\left(
\begin{array}{cc}
\displaystyle w\!\int\limits^{\,\,x}\!(qr_x+iq^2r^2/2)\,dx&0\\
0&\displaystyle w^{-1}\!\int\limits^{\,\,x}\!(q_xr-iq^2r^2/2)\,dx
\end{array}
\right),
$$
$$
B^{(1)}=\left(
\begin{array}{cc}
0&u\\
v&0
\end{array}
\right),\quad
B^{(2)}=\left(
\begin{array}{cc}
\displaystyle -ix+\!\int\limits^{\,\,x}\!qv\,dx&0\\
0&\displaystyle ix+\!\int\limits^{\,\,x}\!ru\,dx
\end{array}
\right), 
$$
where 
$$
w=\mbox{exp}\Bigl(i\!\int\limits^{\,\,x}\!qr/2\,dx\Bigr),\quad
u=\int\limits^{\,\,x}\!q\,dx,\quad v=\int\limits^{\,\,x}\!r\,dx. 
$$

\section{Darboux transformation and infinitesimal symmetries}
Hierarchy of nonlinear equations  (\ref{h}) follows also from the 
compatibility condition of dual Lax pair
\begin{equation}
\xi_x=-\xi\,U(\symbol{26}),
\label{xi_x}
\end{equation}
\begin{equation}
\xi_t=-\xi\,V(\symbol{26}).
\label{xi_t}
\end{equation}
Here $\xi\equiv\xi(x,t,\symbol{26})$ is a vector-row solution, $\symbol{26}$ 
is the spectral parameter of the dual pair.
Since matrix coefficients $U(\lambda)$ and $V(\lambda)$ defined by 
Eqs.(\ref{U}--\ref{V's}) satisfy conditions 
\begin{equation}
\sigma_1U(-\lambda)+U(\lambda)^T\sigma_1=0,\quad
\sigma_1V(-\lambda)+V(\lambda)^T\sigma_1=0,
\label{cond1}
\end{equation}
where $\sigma_1$ is Pauli matrix
$$
\sigma_1=
\left(
\begin{array}{cc}
0&1\\
1&0
\end{array}
\right),
$$
the next connection between the solutions of systems (\ref{psi_x},\ref{psi_t}) 
and (\ref{xi_x},\ref{xi_t}) exists:
\begin{equation}
\xi=\psi^T\sigma_1,\quad\symbol{26}=-\lambda.
\label{conn1}
\end{equation}
In the case of reduction (\ref{rc}) the solutions with complex conjugate 
spectral parameters are also connected. 
For instance, $(\psi_2^*,\pm\psi_1^*)^T$ is a solution of direct Lax pair 
(\ref{psi_x},\ref{psi_t}) with spectral parameter $\lambda^*$.

Let vector-column $\varphi=(\varphi_1,\varphi_2)^T$ and vector-row 
$\chi=(\chi_1,\chi_2)$ are solutions of Lax pairs (\ref{psi_x},\ref{psi_t}) 
and (\ref{xi_x},\ref{xi_t}) with spectral parameters $\mu$ and $\nu$, 
respectively.
The Lax pairs are covariant with respect to ''turned'' binary Darboux 
transformation (BDT) $\{\psi,\xi,U(\lambda),V(\lambda)\}\to
\{\psi[1],\xi[1],U(\lambda)[1],V(\lambda)[1]\}$ of the form 
\begin{equation}
\psi[1]=gT(\lambda)\psi,\quad
\xi[1]=\xi T(\symbol{26})^{-1}g^{-1},
\label{t_psi_xi}
\end{equation}
\begin{equation}
U(\lambda)[1]=\lambda^2U^{(2)}[1]+\lambda\,U^{(1)}[1],\quad
V(\lambda)[1]=\sum_{j=1}^{2m}\lambda^jV^{(j)}[1],
\label{t_U_V}
\end{equation}
where 
$$
T(\lambda)=E-\frac{\mu-\nu}{\lambda-\nu}P=
\Bigl(1-\frac{\nu-\mu}{\lambda-\mu}P\Bigr)^{-1},\quad
P=\frac{\varphi\chi}{\chi\varphi},\quad g=\sigma_1T(0)^{-1}
$$
and
\begin{equation}
\vphantom{\sum\limits_{k=j+1}^{2m}}
U^{(2)}[1]=gU^{(2)}g^{-1},\quad V^{(2m)}[1]=gV^{(2m)}g^{-1},
\label{t_U2}
\end{equation}
\begin{equation}
\vphantom{\sum\limits_{k=j+1}^{2m}}
U^{(1)}[1]=g\Bigl(U^{(1)}+(\mu-\nu)[U^{(2)},P]\Bigr)g^{-1},
\label{t_U1}
\end{equation}
\begin{equation}
V^{(j)}[1]=gV^{(j)}g^{-1}+(\mu-\nu)\sum\limits_{k=j+1}^{2m}\nu^{k-j-1}
\Bigl(V^{(k)}[1]gP-gP\,V^{(k)}\Bigr)g^{-1}
\label{t_V}
\end{equation}
($j=1,...,2m-1$).
We call this transformation as ''turned'' because formulas 
(\ref{t_psi_xi}--\ref{t_V}) is a product of usual BDT \cite{U_1,LU} and 
additional gauge transformation performed with the help of matrix $g$.
This additional transformation allows us to avoid an appearance of the terms 
at the zero power of $\lambda$ in the expressions for $U(\lambda)[1]$ and 
$V(\lambda)[1]$. 

Conditions (\ref{cond1},\ref{conn1}) are fulfilled for transformed matrix 
coefficients $U(\lambda)[1]$, $V(\lambda)[1]$ and solutions $\psi[1]$, 
$\xi[1]$ of the transformed Lax pairs if we impose restriction 
$$
\chi=\varphi^T\sigma_1,\quad\nu=-\mu.
$$
In this case, we have  
$$
U^{(2)}[1]=U^{(2)},
$$
$$
V^{(2m)}[1]=V^{(2m)}. 
$$ 
Then, Eq.(\ref{t_U1}) gives us expressions for new (transformed) solutions of 
hierarchy of nonlinear equations (\ref{h})\/:
$$
q[1]=r-\frac{1}{\mu}\left(\frac{\varphi_2}{\varphi_1}\right)_{\!x},
$$
$$
r[1]=q-\frac{1}{\mu}\left(\frac{\varphi_1}{\varphi_2}\right)_{\!x}. 
$$
The second iteration of the BDT (\ref{t_psi_xi}--\ref{t_V}) keeping conditions 
(\ref{cond1},\ref{conn1}) yields the following formulas
\begin{equation}
q[2]=q-\frac{\mu_1^2-\mu_2^2}{\mu_1\mu_2}
\left(\frac{\varphi_1^{(1)}\varphi_1^{(2)}}
{\mu_1\varphi_1^{(1)}\varphi_2^{(2)}-\mu_2\varphi_2^{(1)}\varphi_1^{(2)}}
\right)_{\!x},
\label{t_q_2}
\end{equation}
\begin{equation}
r[2]=r+\frac{\mu_1^2-\mu_2^2}{\mu_1\mu_2}
\left(\frac{\varphi_2^{(1)}\varphi_2^{(2)}}
{\mu_2\varphi_1^{(1)}\varphi_2^{(2)}-\mu_1\varphi_2^{(1)}\varphi_1^{(2)}}
\right)_{\!x},
\label{t_r_2}
\end{equation}
where $\varphi^{(k)}_1$ and $\varphi^{(k)}_2$ are the components of vector 
solution $\varphi^{(k)}$ of the direct Lax pair with spectral parameters 
$\mu_k$ ($k=1,2$).
If we put here $\varphi^{(2)}=({\varphi^{(1)}_2}^*,\pm{\varphi^{(1)}_1}^*)^T$
and $\mu_2=\mu_1^*$, then 
$$
r[2]=\pm q[2]^*.
$$
This way we come to DT for the DNLSE hierarchy.
The compact form of $N$-th iteration of this transformation is presented in
\cite{St_3}.

Considering limits $\mu_1\to\mu$ and $\mu_2\to\mu$ in 
Eqs.(\ref{t_q_2},\ref{t_r_2}), one obtains the next expressions (up to a 
multiplier) for solution of the linearization of system (\ref{h}): 
\begin{equation}
\delta q=\left(\varphi_1^{(1)}\varphi_1^{(2)}\right)_{\!x},
\label{iq}
\end{equation}
\begin{equation}
\delta r=-\left(\varphi_2^{(1)}\varphi_2^{(2)}\right)_{\!x}.
\label{ir}
\end{equation}
It is checked by straightforward calculation that 
\begin{equation}
R\left(\begin{array}{c}\delta q\\ \delta r\end{array}\right)
=\mu^2\left(\begin{array}{c}\delta q\\ \delta r\end{array}\right).
\label{Rd}
\end{equation}
This identity allows us to define in a recurrent manner the coefficients of 
expansions of the right--hand sides of Eqs.(\ref{iq},\ref{ir}) in the power 
series on the spectral parameter at a neighborhood of the points $\mu=\infty$ 
and $\mu=0$. 
The coefficients of these expansions 
$$
\delta q=\sum\limits_{k=0}^{\infty}\mu^{-2k}\delta q^{(k)},\quad
\delta r=\sum\limits_{k=0}^{\infty}\mu^{-2k}\delta r^{(k)}
$$
and 
$$
\delta q_j=\sum\limits_{k=0}^{\infty}\mu^{2k}\delta q_j^{(k)},\quad
\delta r_j=\sum\limits_{k=0}^{\infty}\mu^{2k}\delta r_j^{(k)}
$$
form the infinite hierarchies of infinitesimal symmetries.
Operator $R$ satisfying (\ref{Rd}) is nothing but the recursion operator of 
the hierarchy (\ref{h}). 
In the case of point $\mu=0$, there exist three hierarchies of nonlocal
infinitesimal symmetries $\delta q_j^{(k)}$, $\delta r_j^{(k)}$ 
($j=1,2,3$, $k=0,1,...$) that correspond to different choices of the constants 
of integration in operator $R^{-1}$ (see Eq.(\ref{invR})).
The first nontrivial members of the hierarchies for the points $\mu=\infty$ 
and $\mu=0$, respectively, are 
$$
\left\{
\begin{array}{l}
\delta q^{(1)}=q_x\\
\delta r^{(1)}=r_x
\end{array}
\right.\!,\quad
\left\{
\begin{array}{l}
\delta q^{(2)}=q_t/2\\
\delta r^{(2)}=r_t/2
\end{array}
\right.\!,\quad
\left\{
\begin{array}{l}
\delta q^{(3)}=(-q_{xx}+3iq_xqr+3q^3r^2/2)_x/4\\
\delta r^{(3)}=(-r_{xx}-3ir_xqr+3q^2r^3/2)_x/4
\end{array}
\right.\!,
$$
$$
\left\{
\begin{array}{l}
\delta q_1^{(1)}=q\\
\delta r_1^{(1)}=-r
\end{array}
\right.\!,\quad
\left\{
\begin{array}{l}
\delta q_2^{(1)}=2(qv-i)\\
\delta r_2^{(1)}=-2rv
\end{array}
\right.\!,\quad
\left\{
\begin{array}{l}
\delta q_3^{(1)}=2qu\\
\delta r_3^{(1)}=-2(ru+i)
\end{array}
\right.\!.
$$
It is seen from these formulas that $\delta q^{(k)}\sim v_{12,x}^{(2m-2k+1)}$,
$\delta r^{(k)}\sim v_{21,x}^{(2m-2k+1)}$ and the infinite hierarchy 
corresponding to the point $\mu=\infty$ is local.
Another way of producing the hierarchies of the infinitesimal symmetries is to 
substitute expansions (\ref{psi_inf}) and (\ref{psi_0}) into 
Eqs.(\ref{iq},\ref{ir}).

\section{Conservation laws and Noether's theorem}
Let us consider identity 
$$
(\xi\psi)_{xt}=(\xi\psi)_{tx}. 
$$
Excluding the derivatives of $\psi$ and $\xi$ on $x$ in the left-hand side 
and the derivatives on $t$ in the right-hand side with the help of 
Eqs.(\ref{psi_x},\ref{psi_t}) and (\ref{xi_x},\ref{xi_t}), respectively, 
and dividing the relation obtained on $\lambda-\symbol{26}$, we come to the 
conservation law of the DNLSE hierarchy
$$
T_t+X_x=0,
$$
where
\begin{equation}
T=\xi\Bigl((\lambda+\symbol{26})U^{(2)}+U^{(1)}\Bigr)\psi,
\label{T}
\end{equation}
\begin{equation}
X=-\xi\sum_{k=1}^{2m}\sum_{j=0}^{k-1}\lambda^{k-j-1}\symbol{26}^jV^{(k)}\psi.
\label{X}
\end{equation}
If we put $\lambda=\symbol{26}=\mu$, $\psi=\varphi^{(1)}$ and 
$\xi=(\varphi_2^{(2)},-\varphi_1^{(2)})$, where vectors 
$\varphi^{(k)}=(\varphi_1^{(k)},\varphi_2^{(k)})^T$ ($k=1,2$), as it was 
supposed at the end of the previous section, are solutions of Lax pair 
(\ref{psi_x},\ref{psi_t}) with spectral parameter $\mu$, then expressions 
(\ref{T},\ref{X}) are rewritten in the next manner
\begin{equation}
T=-2i\mu(\varphi_1^{(1)}\varphi_2^{(2)}+\varphi_2^{(1)}\varphi_1^{(2)})
+q\varphi_2^{(1)}\varphi_2^{(2)}-r\varphi_1^{(1)}\varphi_1^{(2)},
\label{T_phi}
\end{equation}
\begin{equation}
\begin{array}{c}
\displaystyle X=-2\sum_{k=1}^{m}k\mu^{2k-1}v^{(2k)}
(\varphi_1^{(1)}\varphi_2^{(2)}+\varphi_2^{(1)}\varphi_1^{(2)})+{}\\
\displaystyle{}+\sum_{k=1}^{m}(2k-1)\mu^{2k-2}
\left(v_{21}^{(2k-1)}\varphi_1^{(1)}\varphi_1^{(2)}-
v_{12}^{(2k-1)}\varphi_2^{(1)}\varphi_2^{(2)}\right).
\end{array}
\label{X_phi}
\end{equation}
Substitution of expansions (\ref{psi_inf}) and (\ref{psi_0}) of the Lax pair 
solutions at the neighborhood of points $\mu=\infty$ and $\mu=0$ into these 
formulas leads to the hierarchies of the conservation laws expressed 
explicitly through the solutions of nonlinear equations (\ref{h}).
In the case of point $\mu=0$, for example, we have three infinite hierarchies 
$T_{j,t}^{(k)}+X_{j,x}^{(k)}=0$ ($j=1,2,3$, $k=0,1,...$), whose first 
conserved densities and currents are 
\begin{equation}
T_1^{(0)}=q,\quad X_1^{(0)}=-v_{12}^{(1)},\quad
T_2^{(0)}=r,\quad X_2^{(0)}=-v_{21}^{(1)},
\label{TX_0a}
\end{equation}
\begin{equation}
T_3^{(1)}=qv-ur,\quad X_3^{(1)}=uv_{21}^{(1)}-vv_{12}^{(1)}-2v^{(2)}.
\label{TX_0b}
\end{equation}
The first two conservation laws are immediate consequence of the divergent 
form of Eqs.(\ref{h}).

Let us discuss the connection between solutions (\ref{iq},\ref{ir}) of the 
linearized equations and the conservation laws found in the case of system of 
nonlinear equations
\begin{equation}
iq_t+q_{xx}-i(q^2r)_x=0,
\label{q_t}
\end{equation}
\begin{equation}
ir_t-r_{xx}-i(qr^2)_x=0.
\label{r_t}
\end{equation}
Coefficients of the second equation of Lax pair (\ref{psi_x},\ref{psi_t}) of 
the system under consideration are 
$$
V^{(4)}=2U^{(2)},\quad V^{(3)}=2U^{(1)},\quad V^{(2)}=qrU^{(2)},\quad
V^{(1)}=
\left(
\begin{array}{cc}
0&iq_x+q^2r\\
-ir_x+qr^2&0
\end{array}
\right)
$$
DNLSE (\ref{DNLSE}) follows these equations by imposing condition $r=\pm q^*$. 

In the terms of potentials $u$ and $v$ the Lagrangian of 
Eqs.(\ref{q_t},\ref{r_t}) reads as
$$
{\cal L}=i(u_xv_t+v_xu_t)+u_{xx}v_x-v_{xx}u_x-iu_x^2v_x^2\,.
$$
Using notations for the Euler--Lagrange equations
$$
\Lambda(u)\equiv-\left(\frac{\partial\cal L}{\partial u_t}\right)_t
-\left(\frac{\partial\cal L}{\partial u_x}\right)_x
+\left(\frac{\partial\cal L}{\partial u_{xx}}\right)_{xx}=
-2(ir_t-r_{xx}-i(qr^2)_x)=0,
$$
$$
\Lambda(v)\equiv-\left(\frac{\partial\cal L}{\partial v_t}\right)_t
-\left(\frac{\partial\cal L}{\partial v_x}\right)_x
+\left(\frac{\partial\cal L}{\partial v_{xx}}\right)_{xx}=
-2(iq_t+q_{xx}-i(q^2r)_x)=0,
$$
the variation of the Lagrangian, which is caused by the infinitesimal 
transformations of potentials $u\to u+\varepsilon\delta u$ and 
$v\to v+\varepsilon\delta v$, is written in a form of Noether's identity
\begin{equation}
\delta{\cal L}=\varepsilon(A_t+B_x+\Lambda(u)\delta u+\Lambda(v)\delta v).
\label{Ni}
\end{equation}
Here
$$
A=\frac{\partial\cal L}{\partial u_t}\,\delta u+
\frac{\partial\cal L}{\partial v_t}\,\delta v=i(q\,\delta v+r\,\delta u),
$$
$$
B=\left(\frac{\partial\cal L}{\partial u_x}-
\left(\frac{\partial\cal L}{\partial u_{xx}}\right)_x\,\right)\delta u+
\frac{\partial\cal L}{\partial u_{xx}}\,\delta u_x+
\left(\frac{\partial\cal L}{\partial v_x}-
\left(\frac{\partial\cal L}{\partial v_{xx}}\right)_x\,\right)\delta v+
\frac{\partial\cal L}{\partial v_{xx}}\,\delta v_x=
$$
$$
=q_x\,\delta v-r_x\,\delta u+r\,\delta q-q\,\delta r-iqr^2\,\delta u-
iq^2r\,\delta v.
$$
Given a symmetry of Eqs.(\ref{q_t},\ref{r_t}), a conservation law is derived 
from Eq.(\ref{Ni}) due to the Noether's theorem. 
Few examples of the symmetries and associated conservation densities and
currents are listed below:

\noindent{\bf 1)} $\delta u=1$, $\delta v=0$:
$$
\delta {\cal L}=0,
$$
\begin{equation}
T_1=ir,\quad X_1=iv_t-2r_x-2iqr^2.
\label{TX_1}
\end{equation}
\noindent{\bf 2)} $\delta u=0$, $\delta v=1$:
$$
\delta {\cal L}=0,
$$
\begin{equation}
T_2=iq,\quad X_2=iu_t+2q_x-2iq^2r.
\label{TX_2}
\end{equation}
\noindent{\bf 3)} $u\to u\mbox{e}^{i\varepsilon}$, 
$v\to v\mbox{e}^{-i\varepsilon}$, $\delta u=iu$, $\delta v=-iv$:
$$
\delta {\cal L}=0,
$$
\begin{equation}
T_3=qv-ur,\quad X_3=u_tv-uv_t-2i(q_xv-qr+ur_x)-2(qv-ur)qr.
\label{TX_3}
\end{equation}
\noindent{\bf 4)} $x\to x+\varepsilon$, $\delta u=q$, $\delta v=r$:
$$
\delta {\cal L}=\varepsilon{\cal L}_x,
$$
\begin{equation}
T_4=2iqr,\quad X_4=2(q_xr-qr_x)-3iq^2r^2.
\label{TX_4}
\end{equation}
\noindent{\bf 5)} $t\to t+\varepsilon$, $\delta u=u_t$, $\delta v=v_t$:
$$
\delta {\cal L}=\varepsilon{\cal L}_t,
$$
\begin{equation}
T_5=-q_xr+qr_x+iq^2r^2,\quad X_5=i(q_{xx}r-2q_xr_x+qr_{xx}-2q^3r^3)+
3(q_xr-qr_x)qr.
\label{TX_5}
\end{equation}
Conservation laws that arise in the first and second cases are trivial.
The symmetries of the potentials in the third, fourth and fifth cases 
correspond, respectively, to infinitesimal symmetries $\delta q_1^{(1)}$, 
$\delta r_1^{(1)}$, $\delta q^{(1)}$, $\delta r^{(1)}$ and $\delta q^{(2)}$, 
$\delta r^{(2)}$ presented at the end of previous section.
Conserved density $T_5$ is proportional to the Hamiltonian density of DNLSE 
\cite{GIK}.
The Noether's theorem was applied in \cite{DF} to obtain $T_3$ and $X_3$,
which are nothing but $T_3^{(1)}$ and $X_3^{(1)}$ (\ref{TX_0b}). 
Hence, $\delta r_1^{(1)}$ and $\delta q^{(1)}$ are connected by the Noether's 
theorem with $T_3^{(1)}$ and $X_3^{(1)}$. 
It will be proven in the sequel that this is valid for all members of the 
hierarchies of infinitesimal symmetries and conservation laws. 

Formulas (\ref{iq},\ref{ir}) give us solutions of the linearized equations on
potentials
$$
\delta u=\varphi_1^{(1)}\varphi_1^{(2)},
$$
$$
\delta v=-\varphi_2^{(1)}\varphi_2^{(2)}.
$$
It is remarkable that we are able to put the corresponding variation of 
Lagrangian in divergent form: 
$$
\delta{\cal L}=\left(iu\,\delta r+iv\,\delta q+
4\mu(\varphi_1^{(1)}\varphi_2^{(2)}+\varphi_2^{(1)}\varphi_1^{(2)})\right)_t+
$$
$$
+\left(\vphantom{\varphi_1^{(1)}}
v\,\delta q_{{}\,x}-u\,\delta r_{{}\,x}+q\,\delta r-r\,\delta q-
i(ur+2qv)r\,\delta q-i(qv+2ur)q\,\delta r+
\right.
$$
$$
\left.
+8i\mu^2(r\delta u+q\delta v)
-16\mu^3(\varphi_1^{(1)}\varphi_2^{(2)}+\varphi_2^{(1)}\varphi_1^{(2)})
\right)_x\,.
$$
Combining this expression with Eq.(\ref{Ni}), we come after a cancellation of 
the terms with potentials $u$ and $v$ to the conservation law, whose conserved 
density $\tilde T$ and current $\tilde X$ are defined in the following manner
$$
\tilde T=4\mu(\varphi_1^{(1)}\varphi_2^{(2)}+\varphi_2^{(1)}\varphi_1^{(2)})+
2i(q\varphi_2^{(1)}\varphi_2^{(2)}-r\varphi_1^{(1)}\varphi_1^{(2)}),
$$ 
$$
\tilde X=-(16\mu^3+4\mu qr)
(\varphi_1^{(1)}\varphi_2^{(2)}+\varphi_2^{(1)}\varphi_1^{(2)})+
12i\mu^2(r\varphi_1^{(1)}\varphi_1^{(2)}-q\varphi_2^{(1)}\varphi_2^{(2)})+
$$
$$
+2(r_x+iqr^2)\varphi_1^{(1)}\varphi_1^{(2)}
+2(q_x-iq^2r)\varphi_2^{(1)}\varphi_2^{(2)}).
$$
These expressions are proportional to ones given by 
Eqs.(\ref{T_phi},\ref{X_phi}).
This way, we show that solutions (\ref{iq},\ref{ir}) of the linearized 
equations and conserved densities (\ref{T_phi}) and currents (\ref{X_phi}) are 
connected in the case of DNLSE in accordance with the Noether's theorem.
This connection takes also place between the infinite hierarchies of 
infinitesimal symmetries $\delta q^{(k)}$, $\delta r^{(k)}$ and  
$\delta q_j^{(k)}$, $\delta r_j^{(k)}$ ($j=1,2,3$, $k=0,1,...$) and the 
hierarchies of conservation laws obtained by expansion in formulas 
(\ref{T_phi},\ref{X_phi}) of the Lax pair solutions on the spectral parameter 
powers.
First terms of expansions (\ref{psi_inf}) and (\ref{psi_0}) lead to the 
conservation laws determined by formulas (\ref{TX_4},\ref{TX_5}) and 
(\ref{TX_0a},\ref{TX_0b}), respectively, that coincide with ones presented in 
\cite{GIK,S_1,W}.

\section{Conclusion}
In the present report, we have found the expressions for the solution of the 
linearization of the DNLSE hierarchy equations and their conservation law 
in the terms of the solutions of associated Lax pairs. 
The approach exploited is based on the Darboux transformation technique. 
It is shown in the DNLSE case that the conservation law is connected with the 
solution of the linearized equation accordingly to the Noether's theorem. 
The local hierarchy and three nonlocal ones of the infinitesimal symmetries 
and the conservation laws that are explicitly expressed through the variables 
of the nonlinear equations are produced using the recursion operator and/or 
expanding the Lax pair solutions in the series on the spectral parameter 
powers. 

The explicit form of the infinitesimal symmetries and the conservation laws of 
various hierarchies is useful to determine an integrability of the nonlinear 
PDE's given. 
This is especially important for the cases interesting from the physical point 
of view, such as the reductions of the PDE's and their deformations. 
Recently, it was revealed that some deformations of the well-known nonlinear 
integrable equations, which have the physical meaning, are also integrable 
\cite{Z1,Z2}. 
This opens the problems of a description of the classes of the deformations 
keeping the integrability and an extension to them of the methods having been 
developed in the IST theory. 
The approach suggested here is not specific for the hierarchy considered and 
can be applied to other integrable hierarchies and their integrable 
deformations. 
An investigation of the hierarchy of the deformed nonlinear equations, which 
is associated with the quadratic bundle and contains as a particular case the 
following integrable deformation of the DNLSE equation 
$$
iq_t+\alpha q_x^*+q_{xx}\pm i(|q|^2q)_x=0,
$$
where $\alpha$ is an arbitrary parameter, is a subject of the future work. 

\section{ Acknowledgements }
I am grateful Dr. Heinz Steudel for stimulating discussions and hospitality.
I thank Gottlieb Daimler- und Karl Benz-Stiftung for financial support.

\vfill
\eject
\end{document}